\documentclass[aps,prb,preprint,groupedaddress,nofootinbib]{revtex4-2}
\usepackage{amsmath}
\usepackage{xcolor}
\usepackage{graphicx}
\usepackage[normalem]{ulem}

\begin{document}

\title{Analytical modeling of time-varying and dispersive metasurfaces with surface susceptibility operators}

\author{Suat Barış İplikçioğlu}
\email[]{siplikcioglu15@ku.edu.tr}
\affiliation{Department of Electrical and Electronics Engineering,\\Koç University, Istanbul, Turkey}
\affiliation{Profen Technologies, Istanbul, Turkey}

\author{M. I. Aksun}
\email[]{irsadi.aksun@ozyegin.edu.tr}
\affiliation{Department of Electrical and Electronics Engineering,\\Özyeğin University, Istanbul, Turkey}

\begin{abstract}
With the advent of new fabrication technologies, time-varying metasurfaces have emerged as novel platforms for exotic waveform shaping in microwaves and optics, providing an additional degree of freedom to design dynamically controllable and reconfigurable scatterers. Nevertheless, inherent structural properties and material dispersion significantly complicates the computational analysis of such time-varying structures, highlighting the importance of efficient and intuitive analytical approaches that are generalizable to a wide class of meta-atom topologies and temporal modulation profiles. In this work, surface susceptibilities under generalized sheet transition conditions (GSTCs), widely used to model metasurfaces, are extended to a compact and tractable operator form for the frequency-domain analysis of metasurfaces with arbitrary temporal modulation and material dispersion. By virtue of this formalism, we derive the time-variant versions of simple surface susceptibility formulas for periodic arrays of freestanding and substrated dipolar scatterers through the use of appropriate polarizabilities. The applicability of the proposed approach to the characterization of reflective and transmissive properties of pulsed and Floquet metasurfaces is demonstrated, and verified against full-wave simulations.
\end{abstract}
\maketitle

\pagebreak
\section{Introduction}
Time-varying media are active materials which exhibit dynamic alterations of their constitutive parameters in time scales comparable with optical cycles, and have attained a flurry of interest in microwaves and optics communities in recent years, owing to the exotic properties and capabilities they entail~\cite{Caloz2020,Galiffi2022}. While preliminary studies on materials with time-dependent properties were conducted well before the advent of metamaterials and photonic nanostructures~\cite{Morgenthaler1958,Cassedy1963}, there have been significant theoretical~\cite{ZuritaSanchez2009,Xiao2014,Huidobro2021} and experimental progress~\cite{ReyesAyona2015,Tirole2023,Karimi2023} only in the latest decades. In the meanwhile, novel physics brought in by this effect has influenced the burgeoning field of metasurfaces and flat optics, with applications in frequency translation~\cite{Lee2018,Taravati2021,MorenoRodriguez2024,Stefanini2024}, beam steering~\cite{Karimi2023,Liu2018} and parametric amplification~\cite{Wang2023,Kovalev2024}, among others.

Advances in time-varying metasurfaces and metamaterials have gone hand in hand with novel modeling approaches beyond the usual assumption of linear time-invariance. In addition to full-wave simulation techniques~\cite{Shi2016,Mock2023}, highly-accurate approaches such as rigorous-wave coupled analysis~\cite{Inampudi2018}, T-matrix~\cite{Garg2022} and layer-multiple-scattering method~\cite{Panagiotidis2023} were extended to treat time- and spacetime-variant effects. Owing to the relative complexity of these approaches, simple analytical models such as equivalent circuits~\cite{Wang2020,Kovalev2024,Mostafa2022,MorenoRodriguez2024}, surface impedances/admittances~\cite{Lee2018,Liu2018,Wang2023,Stefanini2024} and bulk spatiotemporal homogenization~\cite{Garg2024} are used in the a priori characterization of dynamic metasurfaces. Among these, surface susceptibilities under generalized sheet transition conditions (GSTCs) are arguably the most versatile analytical modeling approach for metasurfaces: in addition to capturing the intricacies of these structures such as normal polarizabilities and chiral behavior, they are easily generalizable to a variety of meta-atom topologies and largely avert the possible spurious artifacts observed in bulk homogenization techniques~\cite{Achouri2021}. While surface susceptibilities were previously adapted to time-varying regimes, these generally took on the forms of dispersionless time-domain expressions~\cite{Achouri2021}, Floquet expansions of Lorentz oscillator equations~\cite{Tiukuvaara2021} and complex double-time kernels~\cite{Amra2024}, often yielding limited or convoluted formulas with no apparent relationship with geometric and material parameters. It should be noted that efficient analytical expressions for characterization of metasurfaces and thin island films with respect to their constitutive properties (namely the polarizabilities) have already been available~\cite{Kuester2003,Tretyakov2003,Bedeaux2004,Albooyeh2016}, albeit in linear time-invariant form.

In this work, we characterize time-modulated and dispersive metasurfaces through the analytically-derived and easily generalizable approximate surface susceptibilities, directly extending the conventional modeling approaches to a linear time-variant form. To model the response of metasurfaces under  arbitrary temporal modulation (with an external pump wave), material dispersion and realistic physical morphology, we invoke the operator approach by Horsley et al.~\cite{Horsley2023}, which makes the simple analytical frequency-domain formulas available to time-varying materials by virtue of matrix algebra. Under the quasistatic approximation of time-varying scatterer polarizabilities in matrix form, reflection and transmission operators of freestanding and substrated dipolar arrays are extracted in a tractable manner. Unlike the previous work on the homogenization of time-varying metasurfaces~\cite{Garg2024}, the proposed approach retains the complete linear time-variant properties of the metasurface (since the structure is homogenized only spatially) and is not limited to high-modulation frequencies. Drawing from the recent advances in metasurface configurations with pulsed~\cite{Karimi2023} and periodic pumping~\cite{Wang2023}, we demonstrate the accuracy of proposed approach with examples in pulsed and Floquet-modulated metasurfaces, including selected cases pertaining to resonance perturbation by temporal interfaces and Wood anomalies, which agree well with full-wave simulations. 

\section{Theory}
\subsection{Reflection and transmission operators: GSTCs for a dynamic metasurface}
This section introduces the generalized sheet transition operators for a dispersive metasurface, which, in turn, yield the reflection and transmission operators at the interface of the metasurface positioned between two distinct isotropic half-spaces with different dielectric permittivities and magnetic permeabilities. The surrounding media of the metasurface can be either dynamic or static; therefore, for generality, they are represented using constitutive relations in operator form.  By virtue of operator formalism~\cite{Horsley2023}, the time-varying constitutive material parameters, including conductivities, relative permittivities and permeabilities, can be cast into a matrix form within a truncated Fourier basis through discrete Fourier transform (DFT) matrices, which makes the frequency-domain analytical modeling approaches available to linear time-variant electromagnetic systems. As a formative example, we derive the tangential electric and magnetic surface susceptibilities of a dynamic Huygens metasurface that are represented with Lorentz oscillator model:
\begin{align}
	\frac{d^2 \textbf{P}_{\parallel}}{dt^2} + \gamma_{e} f_{\gamma e}(t) \frac{d \textbf{P}_{\parallel}}{dt} + \omega^2_{0e} f_{0e}(t) \textbf{P}_{\parallel}= \varepsilon_0 f_{pe}(t) \omega^2_{pe} L \textbf{E}_{\parallel} \label{eq:lorentzian-model-e}\\
	\frac{d^2 \textbf{M}_{\parallel}}{dt^2} + \gamma_{m} f_{\gamma m}(t) \frac{d \textbf{M}_{\parallel}}{dt} + \omega^2_{0m} f_{0m}(t) \textbf{M}_{\parallel}= f_{pm}(t) \omega^2_{pm} L \textbf{H}_{\parallel}
	\label{eq:lorentzian-model-m}
\end{align}
where $\omega_{p\{e,m\}}$, $\gamma_{\{e,m\}}$ and $\omega_{0\{e,m\}}$ are the equivalent plasma, collision and oscillator resonance frequencies of the Lorentzian metasurface, respectively, and their arbitrary time-dependence $f(t)$ by virtue of an external modulation. The parameter $L$ is the unit thickness (1 nm). Multiplication of time-dependent oscillator strengths and the inbound field quantities on the right handside of Eqs.~\ref{eq:lorentzian-model-e} and~\ref{eq:lorentzian-model-m} can be reformulated into a frequency-domain convolution. Under $e^{-i\omega t}$ time convention, discretization of the field and related densities into the Fourier basis and use of the circular convolution would yield the following matrix expression for the susceptibilities: 
\begin{align}
	\hat{\chi}^{\parallel}_{ee} &=-\omega_{pe}^2 L \left[\omega^2 + i \gamma_{e}  \hat{\mathcal{F}} \bar{f}_{\gamma e}\hat{\mathcal{F}}^{-1} \omega - \omega_{0e}^2 \hat{\mathcal{F}} \bar{f}_{0e} \hat{\mathcal{F}}^{-1}\right]^{-1} \hat{\mathcal{F}} \bar{f}_{pe} \hat{\mathcal{F}}^{-1} \\
	\hat{\chi}^{\parallel}_{mm} &=-\omega_{pm}^2 L \left[\omega^2 + i \gamma_{m}  \hat{\mathcal{F}} \bar{f}_{\gamma m}\hat{\mathcal{F}}^{-1} \omega - \omega_{0m}^2 \hat{\mathcal{F}} \bar{f}_{0m} \hat{\mathcal{F}}^{-1}\right]^{-1} \hat{\mathcal{F}} \bar{f}_{pm} \hat{\mathcal{F}}^{-1}
	\label{eq:lorentz-susceptibility-operator}
\end{align}
where $\mathcal{\hat{F}}$ and $\mathcal{\hat{F}}^{-1}$ are the DFT and inverse DFT matrices, respectively. Furthermore, $\bar{f}_k=\text{diag}[f_k(t)]$ and the hat ($\hat{\phantom{a}}$) denotes a matrix operator. We shall note that, within the context of periodically-modulated systems, resultant constitutive parameters can be regarded as a more compact representation of those obtained from the Floquet expansion method~\cite{Tiukuvaara2021}; furthermore, treatment of aperiodic metasurfaces with this approach can be regarded as periodic in long-period limit. Nevertheless, the operator formalism suggested here yields a more robust and direct matrix representation of the constitutive parameters of the metasurface: since the ensuing expressions are essentially the same as their static counterparts, it is easier to manipulate and gain intuition from them, especially under the presence of material dispersion. In the forthcoming section, it will also be seen that the formulation herein enables a tractable correspondence between individual meta-atom polarizabilities and surface susceptibilities, owing to the close analogy between existing static formulas and their operator forms.

Having defined matrix forms of surface susceptibilities for time-varying and dispersive metasurfaces, we may proceed with the derivation of reflection and transmission operators of a metasurface, positioned on $xy$-plane in between two half-spaces. For the sake of brevity, we will strictly focus on electric susceptibilities under TM-polarization, since all work done here is easily extendable to magnetic susceptibilities and TE-polarization as well. Furthermore, assuming that the meta-atoms are translationally-invariant, the geometry becomes mono-isotropic and, in turn, we are left with only two polarization terms, $P_x$ and $P_z$:
\begin{align}
	P_x = \varepsilon_0 \hat{\chi}_{ee}^{xx} \frac{\left[E_x^1(z=0^-) + E_x^2(z=0^+)\right]}{2} \\
	P_z = \varepsilon_0 \hat{\chi}_{ee}^{zz} \frac{\left[\hat{\varepsilon}_1 E_z^1(z=0^-) + \hat{\varepsilon}_2 E_z^2(z=0^+)\right]}{2}
\end{align} 
where $\hat{\chi}_{ee}^{xx}$ and $\hat{\chi}_{ee}^{zz}$ are the tangential and normal electric surface susceptibility operators, respectively. For the sake of generality, no assumptions are made regarding the electrical properties of the surrounding half-spaces, resulting in an operator form for the relevant parameters in these regions. Hence, the magnetic fields in both regions are written as
\begin{align}
	H_y^{1} &= e^{i k_x x + i \hat{k}_{z1} z } + e^{i k_x x - i \hat{k}_{z1} z } \; \hat{r}_p, \;\;\; (z <0) \\
	H_y^{2} & = e^{i k_x x + i \hat{k}_{z2} z } \; \hat{t}_p, \;\;\; (z > 0) 
\end{align} 
where $\hat{k}_{zm}=\left[\hat{k}_m^2 - k_x^2\right]^{\frac{1}{2}}$ and $\hat{k}_m$ is the wavenumber in the medium $m$. Then, the transverse electric fields are obtained as usual, with careful attention to the order of operations, as some parameters are operators and therefore expressed in matrix form:
\begin{align}
	E_x^{1} &= \hat{Z}_1 \left[e^{i k_x x + i \hat{k}_{z1} z } - e^{i k_x x - i \hat{k}_{z1} z } \; \hat{r}_p\right], \;\;\; (z <0) \\
	E_x^{2} & = \hat{Z}_2 \left[e^{i k_x x + i \hat{k}_{z2} z } \; \hat{t}_p\right], \;\;\; (z > 0) 
\end{align} 
where $\hat{Z}_m= \left[\omega \varepsilon_0 \hat{\varepsilon}_m\right]^{-1} \hat{k}_{zm}$ being the wave impedance for each half-space. Using the generalized transition conditions:
\begin{align}
	H_y^{2}-H_y^{1} &= i\omega \varepsilon_0 \hat{\chi}_{ee}^{xx} \frac{1}{2} \left(E_x^{1}+E_x^{2}\right) \\
	E_x^{2}-E_x^{1} &= -\frac{1}{2} \hat{\chi}_{ee}^{zz} \frac{d}{dx} \left(\hat{\varepsilon_1} E_z^1+ \hat{\varepsilon_2}E_z^2\right)
\end{align}
and a series of algebraic manipulations, expressions relating transmission and reflection operators are obtained at $z=0$ as
\begin{align}
	\hat{t}_p-\hat{r}_p-I &= i\omega \varepsilon_0 \hat{\chi}_{ee}^{xx} \frac{1}{2} \left( \left[\hat{Z}_1 (I-\hat{r}_p) + \hat{Z}_2 \hat{t}_p\right] \right) \\
	\hat{Z}_2\hat{t}_p-\hat{Z}_1\left(I-\hat{r}_p\right) &=\frac{i}{2}\hat{\chi}_{ee}^{zz} \varepsilon_0^{-1}\omega^{-1} k_x^2 \left(I+\hat{r}_p+\hat{t}_p\right)
\end{align}
Then, these expressions can be conveniently cast into a system of linear equations, whose solutions are the reflection and transmission operators:
\begin{equation}
	\begin{bmatrix}
		\hat{X} \hat{Z}_1 - I & I - \hat{X} \hat{Z}_2\\
		\hat{Z}_1 - \hat{Y} & \hat{Z}_2 - \hat{Y}\end{bmatrix}
	\begin{bmatrix}
		\hat{r}_p \\
		\hat{t}_p\end{bmatrix}=
	\begin{bmatrix}
		\hat{X} \hat{Z}_1 + I \\
		\hat{Z}_1 + \hat{Y}\end{bmatrix}
\end{equation}
with $\hat{X}=i \frac{1}{2} \omega\varepsilon_0\hat{\chi}_{ee}^{xx}$ and $\hat{Y}=i \frac{1}{2} \hat{\chi}_{ee}^{zz} \varepsilon_0^{-1}\omega^{-1} k_x^2$. If the metasurface is operating strictly in the paraxial regime and normal polarizabilities of the meta-atoms are weak, $\hat{\chi}_{ee}^{zz}$ can be neglected~\cite{Achouri2021}. This yields simplified expressions for the reflection and transmission operators:
\begin{align}
	\hat{r}_p&= \left[\hat{Z}_1 +\hat{Z}_2 - 2 \hat{Z}_2 \hat{X} \hat{Z}_1 \right]^{-1} \left[\hat{Z}_1 - \hat{Z}_2 - 2 \hat{Z}_2 \hat{X} \hat{Z}_1\right] \\
	\hat{t}_p &= \hat{Z}_2^{-1} \hat{Z}_1 (I-\hat{r}_p)
\end{align}
It can be seen that these expressions reduce to Fresnel equations for vanishing surface susceptibilities. For free-standing metasurfaces $(\hat{Z}_1=\hat{Z}_2)$, the expressions further simplify to:
\begin{align}
	\hat{r}_p&= \left[\hat{X} \hat{Z}_1 - I\right]^{-1} \left[\hat{X} \hat{Z}_1\right] \\
	\hat{t}_p &= I-\hat{r}_p
\end{align}
One of the most common applications of GSTCs is susceptibility retrieval from reflection and transmission coefficients. Inversion formulas for electric and magnetic tangential surface susceptibilities under TE and TM excitation are provided in the supplementary material~\cite{Supplementary}. Coupled with broadband simulation approaches specializing on time-varying media~\cite{Mock2023}, these formulas can be used to efficiently homogenize Floquet metasurfaces, for which the number of harmonics is at a reasonable level for computation. 

\subsection{Approximate analytical formulas for surface susceptibilities}
\begin{figure}[h]
	\includegraphics[width=12cm]{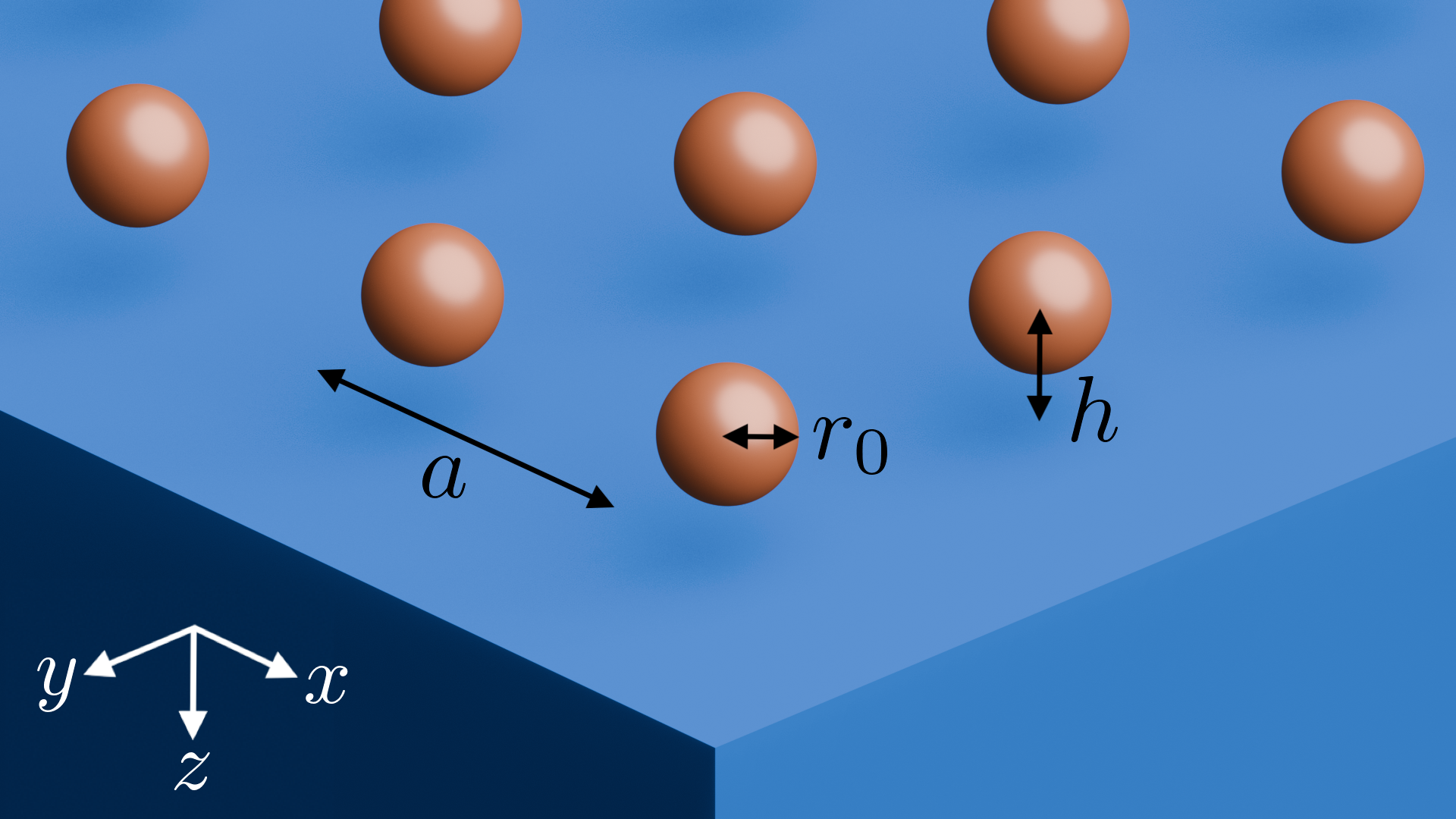}
	\centering
	\caption{A time-varying metasurface, consisting of 2D array of dynamically-modulated spherical scatterers on a substrate.}
	\label{fig:metasurface_substrated}
\end{figure}

Having introduced the reflection and transmission operators for a time-varying surface susceptibilities, we now proceed with the approximate analytical formulas that would represent the physical metasurface as a uniform homogenous sheet. For the sake of illustration, consider a collocation of 2D periodic spheres with a time-varying dielectric constant (Fig. \ref{fig:metasurface_substrated}), with $a$ and $r_0$ being the lattice constant and sphere radius, respectively. Spherical scatterers are chosen as the meta-atoms as the analytical description of their polarizabilities is widely available~\cite{Bedeaux2004}.

For simplification, the meta-atoms are assumed to be in vacuum, with a large lattice spacing and a small size parameter for all frequencies considered. These conditions enable the use of approximate expressions for the polarizabilities~\cite{Tretyakov2003}. Since only the electric dipole term is dominant for these dimensions, the magnetic surface susceptibility can be considered negligible. In addition, the magneto-electric or electro-magnetic sheet susceptibilities are neglected. These parameters particularly influence the response of optically conductive plasmonic scatterers or substrates—an effect known as the substrate-induced bianisotropy (SIB)~\cite{Albooyeh2015}. Even though common time-varying materials such as indium tin oxide (ITO) are optically conductive, we can justify this omission by noting that the nanospheres in quasistatic limit are expected to be smaller than the skin depth of these materials for the incident and induced frequency spectra; this is expected to reduce the effects of SIB~\cite{Albooyeh2015}. Nonetheless, this approximation is made solely for simplicity and is not a limitation of the proposed approach. Thus, we will consider only the tangential and normal electric susceptibility terms $\hat{\chi}^{xx,yy}_{ee}$ and $\hat{\chi}^{zz}_{ee}$; these can be related to the respective dipole polarizabilities of the individual scatterers ($\hat{\alpha}_{ee}^{xx,yy}$ and $\hat{\alpha}_{ee}^{zz}$) using the approximate formulas for tangential and normal surface susceptibilities~\cite{Wind1989} in operator form:
\begin{align}
	\hat{\chi}_{ee}^{xx,yy} &= \left[I- \frac{1}{8\pi a} \left[G_0+\hat{C} G_1\left(\frac{2h}{a}\right)\right] N_a \hat{\alpha}_{ee}^{xx,yy}\right]^{-1}  N_a \hat{\alpha}_{ee}^{xx,yy}
	\label{eq:susceptibility-operator-1} \\
	\hat{\chi}_{ee}^{zz} &= \left[I+ \frac{1}{4\pi a} \left[G_0-\hat{C} G_1\left(\frac{2h}{a}\right)\right] N_a \hat{\alpha}_{ee}^{zz}\right]^{-1}  N_a \hat{\alpha}_{ee}^{zz}
\end{align}
where $N_a=1/a^{2}$ is the density of scatterers and $\hat{C}$ is the image dipole strength, latter of which is scalar for dispersionless and time-invariant substrates. Local field effects due to surrounding scatterers are represented with the term $G_0=9.0336$; in turn, the lattice sum $G_1(x)$ gathers the effects of image dipoles located within the substrate, and is denoted as~\cite{Bagchi1982}:
\begin{equation}
	G(x)=16 \pi^2 \sum_{m=0}^{\infty} \sum_{n=1}^{\infty} \sqrt{m^2+n^2} \; e^{-(2\pi x)\sqrt{m^2+n^2}}
\end{equation}
For free-standing metasurfaces $(\hat{C}=0 \; \text{and} \; h=0)$, these expressions essentially reduce down to the operator forms of sparse approximation formulas by Kuester et al.~\cite{Kuester2003}. Through the use of this quasistatic approximation of the interparticle interactions, which itself is innately intertwined with the spatial homogenization of the metasurface~\cite{Kuester2003}, we refrain from computing cumbersome Sommerfeld integrals or dynamic lattice sums~\cite{Kinayman1995} for each frequency component, at the expense of operating at subwavelength regimes and sparsely-populated array topologies (a valid assumption for many optical metasurface designs). Furthermore, through the use of these frequency-nondispersive interaction fields, we largely avoid the acausal effects that may be observed in the dipolar homogenization formulas for metamaterials~\cite{Alu2011}. 

To model the response of an individual scatter, we will use the modified polarizability expression with a depolarization factor~\cite{Moroz2009} as
\begin{equation}
	\hat{\alpha}_{ee}^{xx,yy,zz}=\left[(\hat{\alpha}_E^{0})^{-1} -\frac{k_0^2}{4\pi r_0}\right]^{-1}
	\label{eq:polarizability-operator}
\end{equation}
with $\hat{\alpha}_E^{0}$ being the static polarizability of a sphere in operator form~\cite{Kristensson1998}:
\begin{equation}
	\hat{\alpha}_{ee}^{0}=4\pi r_0^3 \left[\hat{\varepsilon}_r-I\right] \left[\hat{\varepsilon}_r+2I \right]^{-1}
\end{equation}
In this expression, permittivity operator $(\hat{\varepsilon}_r)$ of the bulk Drude medium with time-dependent oscillator strength $(f(t))$ takes on the form:
\begin{equation}
	\hat{\varepsilon}_r = \varepsilon_{\infty} I - \text{diag}\left[ \frac{\omega_p^2}{\omega^2 + i \omega \gamma} \right] \hat{\mathcal{F}} \text{diag} \left[f(t)\right]\hat{\mathcal{F}}^{-1}
	\label{eq:drude_model}
\end{equation}
We omit the radiation damping term of the modified long-wavelength approximation~\cite{Moroz2009} in Eq.~\ref{eq:polarizability-operator}, since the presence of the infinite lattice of dipolar scatterers compensates for the radiation resistance~\cite{Tretyakov2003}. Spheroidal particles can be treated in a similar manner through the appropriate choice of polarizability for each Cartesian axes~\cite{Bedeaux2004}. Notwithstanding the fact that exact Mie coefficients were rigorously extended to the Floquet regime by several research groups~\cite{Stefanou2021,Ptitcyn2022}, it is much more convient to use the quasistatic polarizability for this problem, owing to the subwavelength nature of the particle with respect to the expected frequency spectra. It should be also mentioned that while similar polarizabilities were previously used to model time-variant scatterers, these were used specifically in the context of adiabatic Floquet media with negligible material dispersion~\cite{Salary2019}, and homogenization of traveling-wave space-time crystals~\cite{Prudencio2024}. To model the bulk material dispersion under pumping, we have adopted the approach suggested by Solís et al.~\cite{Solis2021b} (Eq.~\ref{eq:drude_model}) instead of the low-density approximation~\cite{Mirmoosa2022}, owing to the suitability of the former to temporal interfaces. A detailed discussion on the both models is present in~\cite{Koutserimpas2024}. We shall stress that these approaches yield different permittivity operators due to the change of ordering of the constituent matrices.

Additional information on the analytical formulas and lattice sums presented herein are provided in the supplementary material~\cite{Supplementary}. 
\section{Numerical examples}
In this section, the operator approach presented in the previous section is applied to a time-varying metasurface with spherical meta-atoms in both pulsed and Floquet modulation regimes. The lattice constant $a$ and sphere radius $r_0$ are taken to be 100 nm and 30 nm, respectively. Free-standing and substrated configurations are considered: the distance between the meta-atoms and the substrates is set to 20 nm whenever the latter is present. The time-variant bulk dielectric function of the meta-atoms is modeled after ITO using the Drude model (Eq.~\ref{eq:drude_model}), with parameters obtained from~\cite{Horsley2023}: $\varepsilon_{\infty}=4.08$, $\omega_p=2.88 $ rad/fs and $\gamma=0.145$ rad/fs. For simplicity, the substrate is assumed to be a dispersionless dielectric with $\varepsilon_2=2.1025$, as in silica. To verify the results, we employ the finite element solver COMSOL's time-domain (FETD) and coupled frequency-domain modules for the pulsed and Floquet-modulation schemes, respectively~\cite{Comsol}; a harmonic balance approach~\cite{Shi2016,CorreasSerrano2018} was used for the latter. Details on the simulations are provided in the supplementary material~\cite{Supplementary}.

\subsection{Pulse- and step-modulated metasurfaces}
\begin{figure}[h]
	\includegraphics[width=15cm]{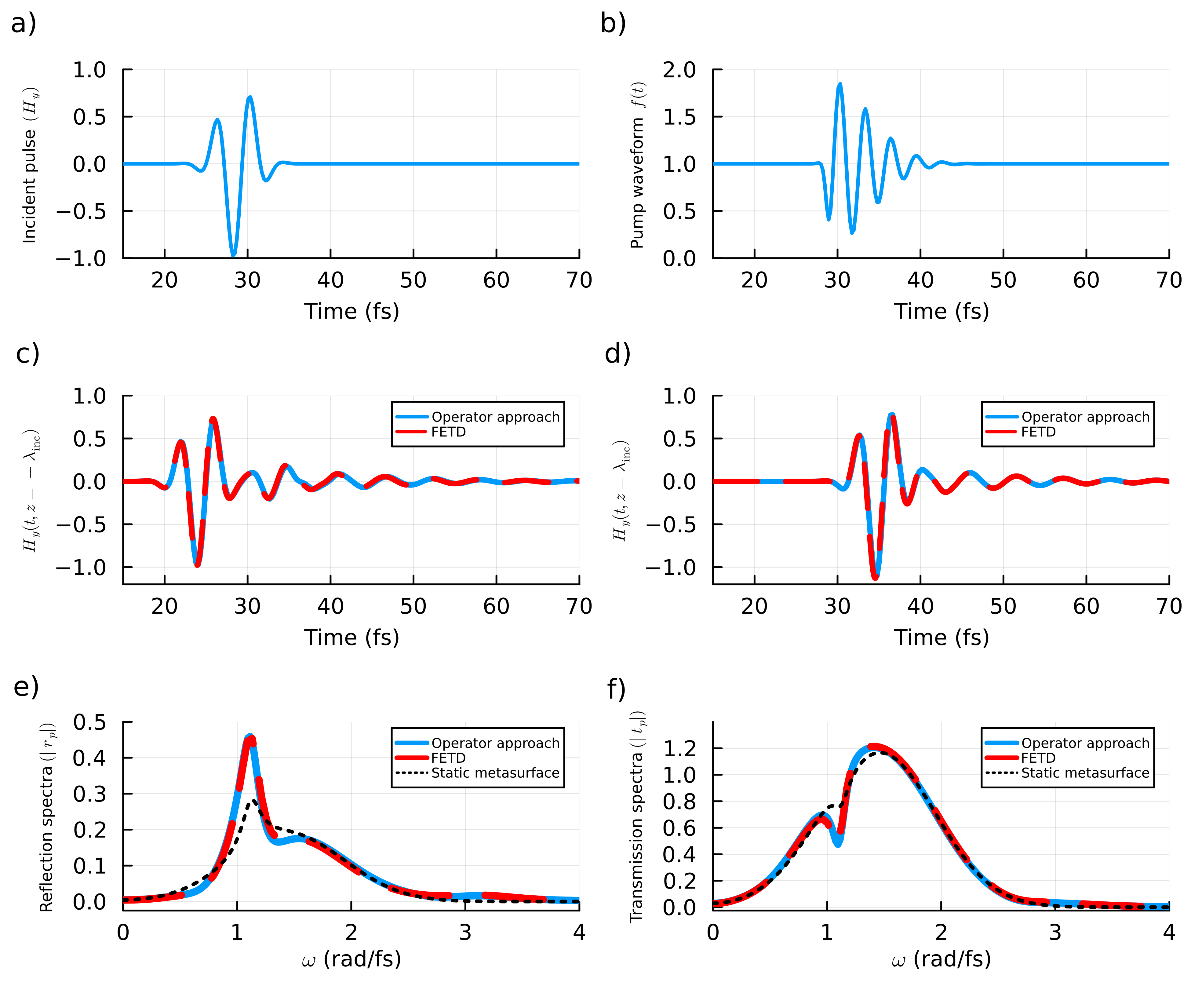}
	\centering
	\caption{Response of a time-varying metasurface under a quasi-periodic pulsed modulation: a) Incident magnetic field probe pulse at $z=-2\lambda_{\text{inc}}$; b) Pump waveform without synchronization; c) Time-domain incident and reflected magnetic field at $z=-\lambda_{\text{inc}}$; d) Time-domain transmitted magnetic field at $z=\lambda_{\text{inc}}$; e) Normalized reflection spectra; f) Normalized transmission spectra. Note that larger-than-unity spectral distribution in (f) is a result of changing wave impedance of the medium.}
	\label{fig:quasiperiodic_pulsed_metasurface_results}
\end{figure}

To highlight the short-time interaction effects in a time-varying metasurface, we study a structure under an arbitrary pulsed modulation of the bulk permittivity of its constituent meta-atoms with an external pump wave. The oscillator strength of the Drude dielectric function is temporally modulated by a quasi-periodic and pulsed waveform with a modulation frequency of $\omega_m$ as the structure is simultaneously excited by a normally-incident $(k_x=0)$, Gaussian-shaped probe wave with a center angular frequency of $\omega_{\text{inc}}=2\pi \cdot 0.23$ rad/fs and a width of 2 fs (Fig.~\ref{fig:quasiperiodic_pulsed_metasurface_results}a). The details of numerical simulations and the modulation waveform — the latter having an approximately decaying sinusoidal shape with $\omega_m=1.4\omega_{\text{inc}}$ and timed such that its peak coincides with the pulse center (Fig.~\ref{fig:quasiperiodic_pulsed_metasurface_results}b)—are provided in the supplementary material~\cite{Supplementary}.

Magnetic field distribution at a single wavelength $(z=\pm \lambda_{\text{inc}} \approx 1300\;\text{nm})$ above and below the metasurface is given in time- and frequency-domain in Fig.~\ref{fig:quasiperiodic_pulsed_metasurface_results}. Near perfect agreement is observed between the results from the operator approach and FETD for the time-domain incident and reflected field profiles, as well as for the transmitted field profile, as shown in Figs.~\ref{fig:quasiperiodic_pulsed_metasurface_results}c and~\ref{fig:quasiperiodic_pulsed_metasurface_results}d, respectively. It can also be observed that both waveforms exhibit an exponential tail, a byproduct of the natural resonance of the metasurface~\cite{Rizza2024}. This resonance can be characterized by a complex frequency $\omega_{\text{res}}' - i\omega_{\text{res}}'' = 1.12 - 0.09i$ rad/fs using the generalized pencil-of-function (GPOF) method~\cite{Hua1989}. As seen in Figs.~\ref{fig:quasiperiodic_pulsed_metasurface_results}e and~\ref{fig:quasiperiodic_pulsed_metasurface_results}f, the resonant peaks occur nearly at the same frequencies for the theoretically calculated values of dynamic and static metasurfaces. Nevertheless, the time-varying metasurface exhibits a much more pronounced resonant reflective peak compared to its unmodulated counterpart, which, in turn, corresponds to a transmission dip at the same frequency. Thus, it can be surmised that the resonant behavior of the metasurface can be modified through the suitable tailoring of the pump waveform, in a similar manner to the approach presented in~\cite{Garg2025}. It should nevertheless be stressed that the resonances observed here is distinct from the plasmonic surface lattice resonances (SLRs)~\cite{Kravets2018}, which are associated with the spatial Rayleigh anomalies beyond the typical homogenization limit of periodic structures. The formulation herein, though easily extendible to dynamic far-field coupling, approximates the interaction constants with a quasistatic lattice sum by virtue of subwavelength periodicity and thus is not suitable for modeling SLRs.
\begin{figure}[h]
	\includegraphics[width=13.5cm]{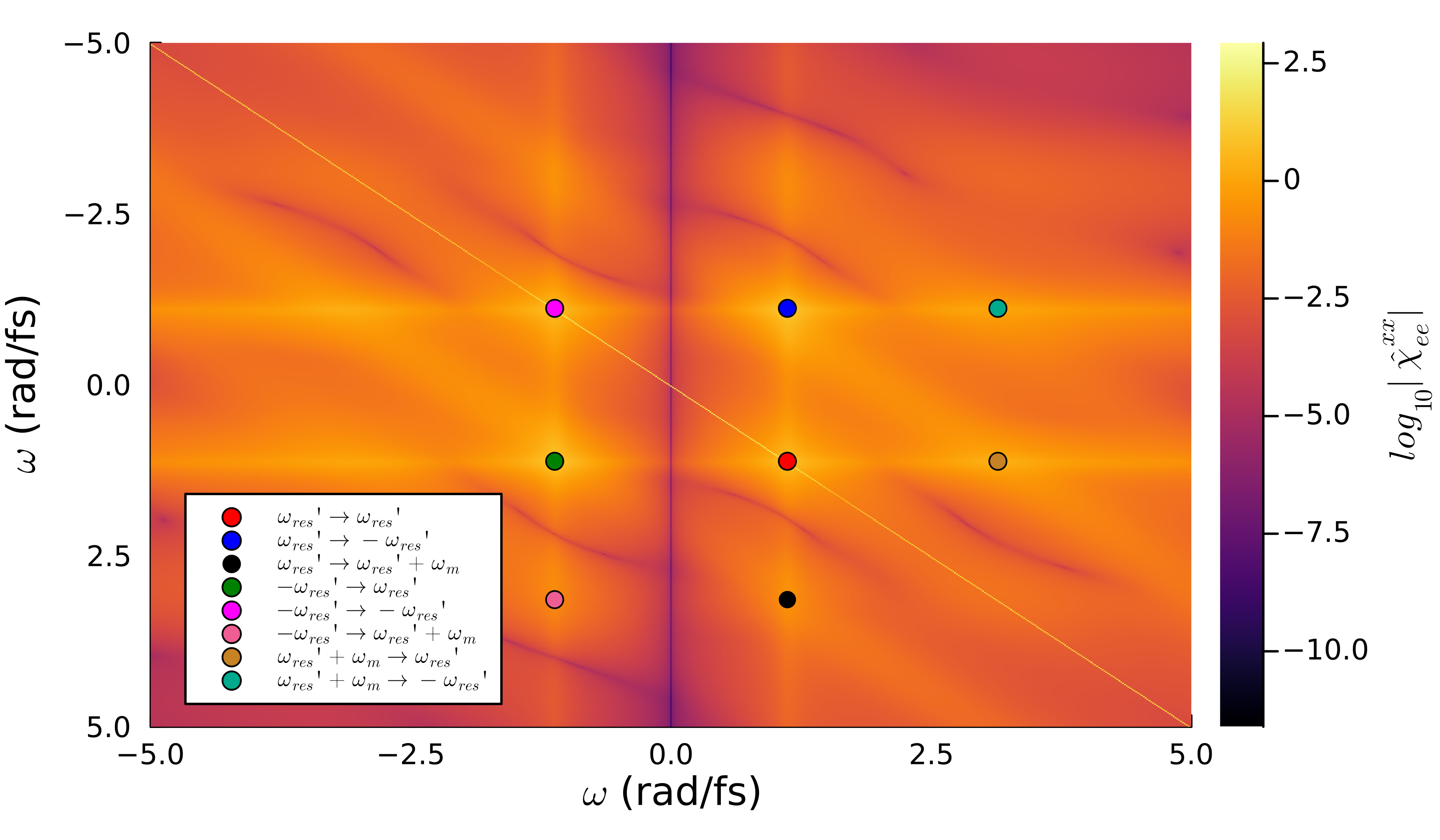}
	\centering
	\caption{Tangential susceptibility spectra of the time-varying metasurface, normalized to 1 nm. Notable coupling frequencies are highlighted.}
	\label{fig:susceptibility_spectra}
\end{figure}

A closer examination of the constitutive operator with modulation but without incident wave reveals a more complex frequency response, which is shown in Fig.~\ref{fig:susceptibility_spectra} as a heatmap of the normalized surface susceptibility operator. As it is expected, there are more resonant frequencies corresponding to quasi-harmonics at $\omega_{\text{res}}'+\omega_m$, though being very weak, in addition to the previously computed resonant frequencies and their intercouplings situated alongside the diagonals, as depicted in Fig.~\ref{fig:susceptibility_spectra}. In addition to the intercoupling between positive and negative resonant frequencies, local peaks at quasi-harmonics can also be observed.

\begin{figure}[h]
	\includegraphics[width=16cm]{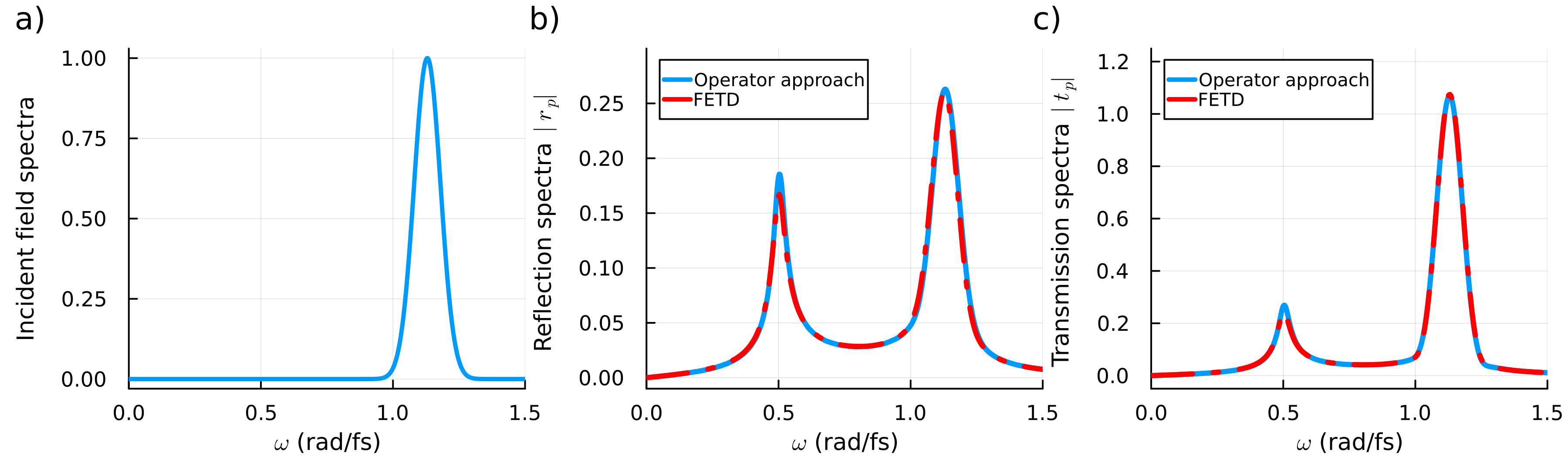}
	\centering
	\caption{Excitation of the perturbed resonant frequency of a substrated metasurface under a temporal interface with a modulation depth of $-0.8$: a) Normalized incident magnetic field spectra; b) Reflected magnetic field spectra and c) Transmitted magnetic field spectra.}
	\label{fig:metasurface_step_spectra}
\end{figure}

Harkening back to Figs~\ref{fig:quasiperiodic_pulsed_metasurface_results}e and~\ref{fig:quasiperiodic_pulsed_metasurface_results}f, we can further observe that the resonant frequency $\omega_{\text{res}}$ is not significantly modified under the pump waveform given on Fig.~\ref{fig:quasiperiodic_pulsed_metasurface_results}b, since the time-dependent oscillator density settles to its original static value well before the excited resonance decays completely, in addition to being relatively slowly-changing. A much stronger effect can be obtained through an abrupt temporal interface in which the oscillator strength increases or decreases in a stepwise manner: the relative change of this parameter with respect to its static value can be defined as the modulation depth. A robust analytical model on the calculation of effective resonant properties of such a temporal interface in the form of a Lorentzian thin-slab with altered plasma frequency was recently introduced by Rizza et al.~\cite{Rizza2024}, who showed the influence of modulation on excitation of guided/radiative modes and off-spectra resonant frequencies. Building on this work, we study the perturbative effects of temporal interfaces on the resonant behavior of the dispersive metasurfaces using the operator formalism. Nevertheless, unlike the thin-slab configuration presented therein, the effective oscillator frequency of a sheet with periodic scatterers is also expected to be modified in addition to the effective plasma frequency: a composite medium with plasmonic inclusions exhibits a Lorentzian behavior, with its effective oscillator frequency being directly proportional to the bulk plasma frequency~\cite{Kristensson1998}. Thus, the temporal variation of the oscillator strength (which directly modifies the plasma frequency) is expected to strongly influence the intrinsic resonant behavior of the structure. This fact has been used to obtain extended momentum bandgaps in photonic time crystals~\cite{Wang2024}. In light of this, we excite the substrated configuration with a normally-incident and narrowband pulse centered at 1.13 rad/fs (close to the resonant frequency of the structure \textit{before} the step change), while a pump wave rapidly and synchronously decreases the oscillator density to 20\% of its original value, corresponding to a modulation depth of $-0.8$. We model this pump wave using a hyperbolic tangent function with a short fall time ($\sim$1.1 fs). Furthermore, in order to accentuate the resonant lineshape, we decrease the collision frequency of the bulk permittivity to a quarter of its original value. Reflection and transmission spectra of the metasurface under step modulation are given on Fig.~\ref{fig:metasurface_step_spectra}. It can be seen that both curves exhibit a sharp peak around 0.5 rad/fs, which corresponds to a resonant frequency of approximately $0.5-0.02i$ rad/fs per GPOF analysis of both analytical and numerical simulation results. As expected from the effective behavior of the metasurface, the induced resonant frequency shifts when the oscillator density rapidly decreases, despite the absence of an overlap with the incident field spectra. Beyond being consistent with the main findings in~\cite{Rizza2024}, this result also highlights the strong perturbative effect of the pumping even for modes with $k_x=0$. In contrast, this effect is absent in the thin-slab configuration, which has an approximate resonant frequency of $\omega_{0e}-i\gamma_e/2$ (per Eq.~\ref{eq:lorentzian-model-e}) for $k_x=0$ independent of the modulation depth. 

\subsection{Floquet metasurfaces}
Having demonstrated the applicability of analytical susceptibility formulas to configurations under pulse- and step-modulation, we now turn our attention to the Floquet metasurfaces under a periodic pumping of the oscillator strength of the bulk dielectric function. We consider a modulation waveform in the form of a cosine function, with a modulation depth and frequency of $0.25$ and $25$ THz, respectively. It should be noted that, unlike more complex modulation waveforms presented in the previous subsection, permittivity operator of this waveform can be obtained through a straightforward Floquet expansion of the Drude differential equation~\cite{Supplementary}. Semi-analytical computations and numerical simulations were performed for up to the fifth harmonic $(n=\pm5)$. 

\begin{figure}[h]
	\includegraphics[width=14cm]{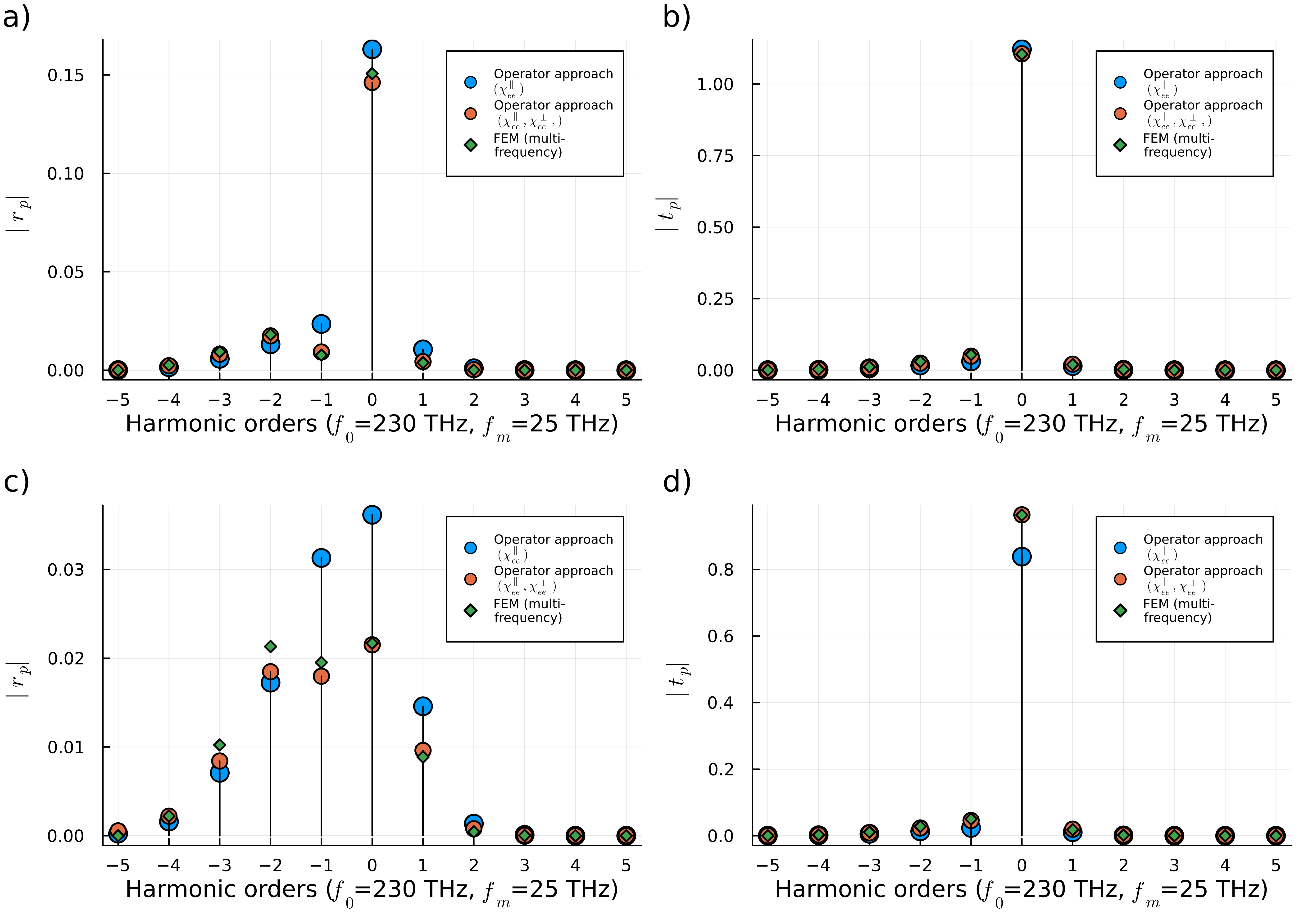}
	\centering
	\caption{Reflection and transmission of Floquet harmonics at $\theta=30^{\circ}$ from a metasurface: a) Reflection coefficient and b) Transmission coefficient for substrated metasurface; c) Reflection coefficient and d) Transmission coefficient for freestanding metasurface.}
	\label{fig:floquet_harmonics_stem_plot}
\end{figure}

Reflection and transmission magnitudes of the substrated and freestanding metasurfaces for the $30^{\circ}$ angle of incidence are given on Fig.~\ref{fig:floquet_harmonics_stem_plot}. For comparison with paraxial approximation of the susceptibilities~\cite{Achouri2021}, reflection and transmission magnitudes are computed both with and without the normal polarization. It is obvious that the operator approach with complete set of polarizabilities results in a good agreement with the full-wave results, while the paraxial approximation particularly overestimates the dominant harmonic $(n=-1)$ for reflection.

\begin{figure}[h]
	\includegraphics[width=14cm]{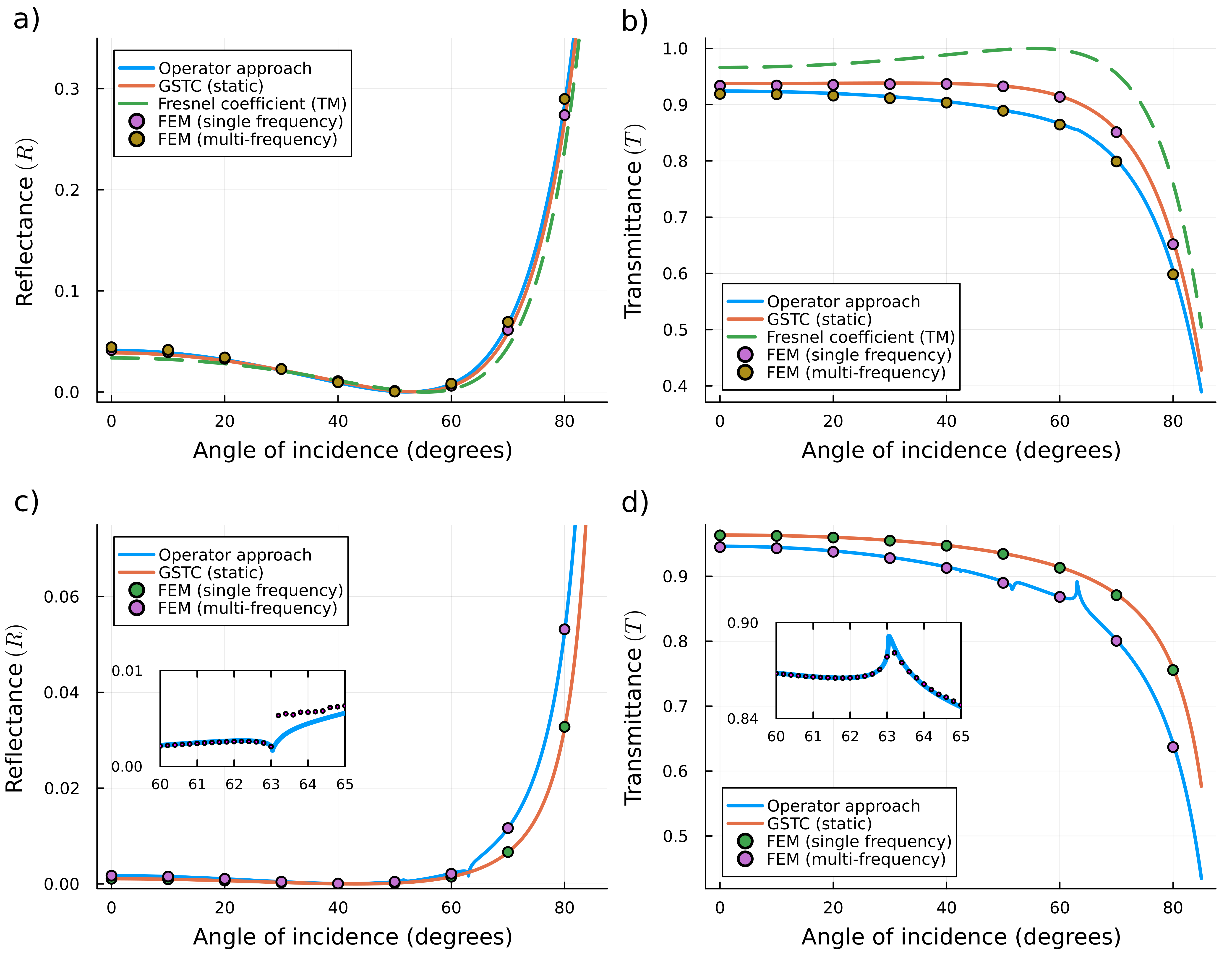}
	\centering
	\caption{Reflectance and transmittance of the Floquet metasurface with respect to different angles of incidence: a) Reflectance and b) Transmittance for substrated metasurface; c) Reflectance and b) Transmittance for freestanding metasurface. For the freestanding metasurfaces, which exhibit more prominent Wood anomalies, a close-up of the $n=-1$ order is provided in the inset.}
	\label{fig:reflectance_transmittance_angle}
\end{figure}

Using the complete set of polarizabilities, the zeroth-order reflectance and transmittance values of the substrated and freestanding metasurfaces with respect to the incidence angle are also computed (Fig.~\ref{fig:reflectance_transmittance_angle}). Close agreement with full-wave simulations can be observed. It can also be seen that perturbation of the reflectance due to temporal modulation is relatively smaller compared to the transmittance, being nearly absent for the case of substrated metasurface. Light transmission under dynamic modulation is observed to be less than that of the static metasurface, which can be attributed to the coupling to nonzero harmonics. Another point to note is the minor resonances at certain angles, which are more visible in the case of freestanding metasurface (Fig.~\ref{fig:reflectance_transmittance_angle}d) and largely supressed with the presence of the substrate (Fig.~\ref{fig:reflectance_transmittance_angle}b). These  essentially correspond to temporal Wood anomalies~\cite{Galiffi2020}: since a wave incident on a periodic temporal interface preserves its transverse wavevector but not its frequency, the resulting idler frequencies that diffract into different angles, with certain negative-order harmonics becoming possibly evanescent. The exact loci of the resonances coincide with the angles in which the excited harmonic becomes evanescent; in analogy with the theory of spatial Wood anomalies~\cite{Hessel1965}, these can be defined as Rayleigh angles and are described with the formula $\theta_n = \sin^{-1} \left(\omega_n/\omega_{\text{inc}}\right)$, where $\omega_n$ is the angular frequency of the  $n$th harmonic. For the given example, these correspond to $27.16^{\circ}$, $34.42^{\circ}$, $42.37^{\circ}$, $51.5^{\circ}$ and $63.04^{\circ}$ for $n=-5,\; -4,\; -3,\; -2, -1$, respectively. As expected, the strongest coupling and the sharpest resonance occurs at $n=-1$. In order to highlight the applicability of surface susceptibilities to temporal Wood anomalies in dipolar arrays of scatterers, closeups of the results around this order are provided at the insets of Figs.~\ref{fig:reflectance_transmittance_angle}c and \ref{fig:reflectance_transmittance_angle}d. While a close level of agreement is observed in general, minor deviations in the reflection coefficient amplitudes at angles larger than the Rayleigh angles are observed (see the inset in Fig.~\ref{fig:reflectance_transmittance_angle}c); these may be attributed to numerical artifacts or emerging dynamic effects. It should be noted that since the operation wavelength and bulk of the excited harmonics of the metasurface are strictly within the homogenization limit, we expect no excitation of spatiotemporal harmonics, which would have led into more complex diffractive behavior~\cite{Hadad2015}.

\section{Conclusion}
In conclusion, we have introduced an analytical methodology for modeling time-varying metasurfaces using surface susceptibility operators, which enables the application of conventional frequency-domain homogenization approaches to linear time-variant systems for obtaining compact, tractable and intuitive formulas. The given surface susceptibilities are able to represent different effects in time-varying photonics such as temporal interfaces and Wood anomalies under material dispersion; the results are in close agreement with full-wave simulations. While the susceptibilities shown here are derived only for electric polarizabilities under a quasistatic approximation, the method is easily extendable to magnetic dipolar and bianisotropic polarizabilities by virtue of Floquet-Mie theory~\cite{Ptitcyn2022} and more intricate interaction constant approaches~\cite{HesariShermeh2021}, respectively, specifically for relatively larger lattice constants and less sparse configurations. Treatment of more complex time-varying particles such as truncated spheres or spheroids on substrates, as well as random scatterers dispersed under a given distribution, is also possible with Bedeaux-Vlieger theory~\cite{Bedeaux2004}.

\section*{Acknowledgement}
The authors thank Ali Mostafazadeh for useful discussions. The authors also acknowledge Koç University Advanced Computing Center for computational resources.

\clearpage
\bibliography{bibliography}

\end{document}